\documentclass[aps,prl,twocolumn,preprintnumbers,showpacs,superscriptaddress,groupedaddress]{revtex4}  

\usepackage{graphicx}  
\usepackage{dcolumn}   
\usepackage{bm}        
\usepackage{amssymb}   
\usepackage{atlasphysics} 
\usepackage{multirow}
\usepackage{subfigure}
\graphicspath{{Figures/}}
\RequirePackage{xspace}
 \usepackage{amsmath}
\usepackage{rotating}
\usepackage{hyperref}

\begin{document}

\preprint{CERN-PH-EP-2011-225}
\preprint{Submitted to Physical Review Letters}
\vspace*{0.2cm}
\title{Observation of a New $\chib$ State in Radiative Transitions to $\Upsilon\mathrm{(1S)}$ and $\Upsilon\mathrm{(2S)}$ at ATLAS}
\author{The ATLAS Collaboration}
\date{December 21, 2011}

\begin{abstract}
\noindent The $\chib(n\textrm{P})$ quarkonium states are produced in proton-proton collisions at the Large Hadron Collider (LHC) at $\sqrt{s} = 7\TeV$ 
and recorded by the ATLAS detector. Using a data sample corresponding to an integrated luminosity of 4.4~\ifb, these states are 
reconstructed through their radiative decays to $\Upsilon\mathrm{(1S, 2S)}$ with $\Upsilon\to\mu^+\mu^-$. In addition to the mass peaks corresponding to the decay modes 
$\chib(1\mathrm{P},2\mathrm{P})\to\Upsilon\mathrm{(1S)}\,\gamma$, a new structure centered at a mass of $10.530\pm 0.005 \stat \pm 0.009 \syst \GeV$ is also observed, 
in both the $\ups1\,\gamma$ and $\ups2\,\gamma$ decay modes. This structure is interpreted as the $\chibp3$ system.
\end{abstract}

\maketitle

Measurements of the properties of heavy quark-antiquark bound states, or
quarkonia, provide a unique insight into the nature of Quantum
Chromodynamics close to the strong decay threshold. For the
$\bbbar$ system, the quarkonium states with parallel quark spins ($s=1$) 
include the S-wave
$\Upsilon$ and the P-wave $\chib$ states, where the latter each comprise
a closely spaced triplet of $J=0,1,2$ spin states: $\chibj0$, $\chibj1$ and
$\chibj2$.  The $\chibp1$ and $\chibp2$, with spin-weighted mass barycenters of 9.90 and
10.26~\GeV, respectively, can be readily produced in the radiative decays of
$\Upsilon(\textrm{2S})$ and $\Upsilon(\textrm{3S})$ and have been studied
experimentally~\cite{ChibObs}. 

In this letter, $\chib$ quarkonium states are reconstructed with the ATLAS detector through the radiative decay modes $\chib{(n\textrm{P})}\to\Upsilon(\textrm{1S})\,\gamma$ and $\chib{(n\textrm{P})}\to\Upsilon{(\textrm{2S})}\,\gamma$, in which $\Upsilon(\textrm{1S, 2S})\to\mu^{+}\mu^{-}$ and the photon is reconstructed either through conversion to $\ee$ or by direct calorimetric measurement.
Previous experiments have measured the $\chibp1$ and $\chibp2$ through these decay modes~\cite{ChiMeas}. The $\chibp3$ state has not previously been observed. 
It is predicted to have an average mass of approximately 10.52~\GeV, with hyperfine mass splitting between the triplet states of 10--20~\MeV~\cite{Chib3P,Chib3P_splitting}.

The ATLAS detector~\cite{ATLAS} is a general-purpose particle physics detector with a forward-backward symmetric cylindrical geometry and near $4\pi$ coverage in solid angle. 
The inner tracking detector (ID) consists of a silicon pixel detector, a silicon microstrip detector, and a transition radiation tracker (TRT). The ID is surrounded by a thin superconducting 
solenoid providing a 2\!~T magnetic field and by high-granularity liquid-argon sampling electromagnetic (EM) calorimeters. An iron-scintillator tile calorimeter provides hadronic coverage 
in the central rapidity range. The endcap and forward regions are instrumented with liquid-argon calorimeters for both electromagnetic and hadronic measurements. 
The muon spectrometer (MS) surrounds the calorimeters and consists of a system of precision tracking chambers and detectors for triggering, inside a toroidal magnetic field.

The data sample used for this measurement was recorded by the ATLAS experiment during the 2011 LHC proton-proton collision run at a center-of-mass energy $\rts=7\TeV$.
The integrated luminosity of the data sample, which includes only data-taking periods where all relevant detector sub-systems were operational, is $4.4\,\ifb$. 
A set of muon triggers designed to select events containing muon pairs or single high transverse momentum muons was used to collect the data sample.

In this analysis each muon candidate must satisfy standard muon quality requirements~\cite{ATLASupsilonxsect}.
It must have a track, reconstructed in the MS, combined with a track reconstructed in the ID with transverse momentum $\pt > 4 \GeV$ and pseudorapidity $|\eta| < 2.3$.
The di-muon selection requires a pair 
of oppositely charged muons, which are fitted to a common vertex. A very loose vertex quality requirement ($\chi^{2}$ per degree of freedom [d.o.f.] $< 20$) is used and no mass or 
momentum constraints are applied to the fit. The di-muon candidate is also required to have $\pt > 12 \GeV$ and $|y| < 2.0$.
The invariant mass distribution, $m_{\mu\mu}$, of di-muon candidates is shown in Fig.~\ref{fig.PaperDiMuonMass}. 
Those candidates with masses in the ranges $9.25 < m_{\mu\mu} < 9.65 \GeV$ and $9.80 < m_{\mu\mu} < 10.10 \GeV$ are 
\begin{figure}[htbp]
\begin{center}
\includegraphics[width=0.5\textwidth]{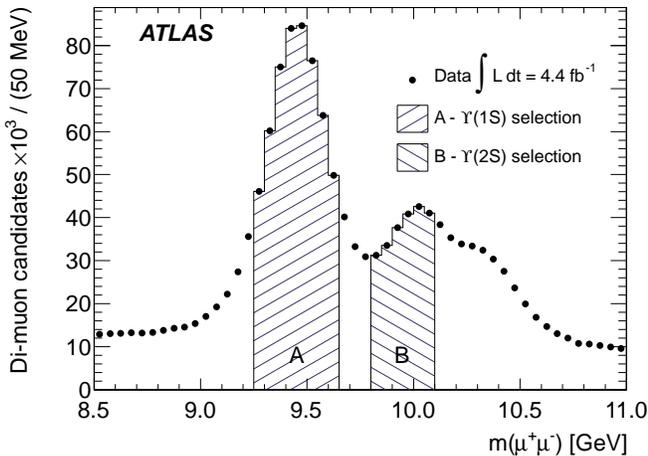}
\caption{The invariant mass of selected di-muon candidates. The shaded regions
 $A$ and $B$ show the selections for $\Upsilon(\textrm{1S})$ and $\Upsilon(\textrm{2S})$ candidates respectively.   }
\label{fig.PaperDiMuonMass}
\end{center}
\end{figure}
selected as $\Upsilon(\textrm{1S})\to\mumu$ and $\Upsilon(\textrm{2S})\to\mumu$ candidates respectively.
The asymmetric mass window (evident from Fig.~\ref{fig.PaperDiMuonMass}) 
for $\Upsilon(\textrm{2S})$ candidates is chosen in order to reduce contamination from the $\Upsilon(\textrm{3S})$ peak and continuum background contributions.

The reconstruction of photons in ATLAS is described in Ref.~\cite{ATLASphotons}. Further details related to this particular analysis are described below.

Converted photons are reconstructed from two oppositely charged ID tracks intersecting at a conversion vertex, with the opening angle between the two tracks at this vertex constrained to be zero.
For tracks with signals in the TRT, the transition radiation should be consistent with an electron hypothesis.
In order to be reliably reconstructed, each conversion electron track must have a minimum transverse momentum of $500 \MeV$.
It is also required to have at least four silicon detector hits and not to be associated to either of the two muon candidates. 
To reduce background contamination, the conversion candidate vertex is required to be at least 40~mm from the beam axis and have a vertex $\chi^{2}$ probability of greater than 0.01. The converted photon impact parameter with respect to the di-muon vertex is required to be less than 2~mm.

Electromagnetic calorimeter energy deposits not matched to any track are classified as unconverted photons.  This analysis uses
the ``loose'' photon selection described in Ref.~\cite{ATLASphotons}, with a minimum photon transverse energy of 2.5 GeV. The loose photon selection includes a limit on the fraction of the 
energy deposit in the hadronic calorimeter 
as well as a requirement that the transverse width of the shower
be consistent with the narrow shape expected for an EM shower. 

To check that an unconverted photon originates from the same vertex as the $\Upsilon$, and to improve the mass resolution of the reconstructed $\chib$, the polar angle 
of the photon is corrected using the procedure described in Ref.~\cite{ATLASHgamgam}. 
The corrected polar angle is determined using the measurement of the photon direction from the longitudinal segmentation of the calorimeter and the constraint from the di-muon vertex position.
Photons incompatible with having originated from the di-muon vertex are rejected by means of a loose cut on the fit result ($\chi^{2}$ per d.o.f. $< 200$). 

The converted (unconverted) photon candidates are required to be within $|\eta| < 2.30\, (2.37)$. Unconverted photons must also be outside the transition region between the barrel and the endcap calorimeters, $1.37 < |\eta| < 1.52$.

The $\chib$ candidates are formed by associating a reconstructed ${\Upsilon}\to\mumu$ candidate with a reconstructed photon. 
The invariant mass difference $\Delta m = m(\mumu \, \gamma)-m(\mumu)$ is calculated to minimize the effect of $\Upsilon\to\mumu$ mass resolution. 
In order to compare the $\Delta m$ distributions of both $\chib(n\textrm{P})\to\Upsilon(\textrm{1S})\gamma$ and $\chib(n\textrm{P})\to\Upsilon(\textrm{2S})\gamma$ decays, 
the variable $\tilde{m}_{k} = \Delta m + m_{\Upsilon(k\textrm{S})}$ is defined, where $m_{\Upsilon(k\textrm{S})}$ are the world average masses~\cite{PDG} of the $\Upsilon(k\textrm{S})$ states.
Requirements of $\pT(\mumu)>20\GeV$ and $\pT(\mumu)>12\GeV$ are applied to $\Upsilon$ candidates with unconverted and converted photon candidates respectively. These thresholds are chosen in order to optimize signal significance in the $\chib(\textrm{1P,2P})$ peaks.

Figure~\ref{fig.PaperChiBMass}(a) shows the $\tilde{m}_{1}$ distribution for unconverted photons and Fig.~\ref{fig.PaperChiBMass}(b) the $\tilde{m}_{1}$ and $\tilde{m}_{2}$
distributions for converted photons. 
\begin{figure}[htbp]
\begin{center}
\subfigure[]{
  \includegraphics[width=0.5\textwidth]{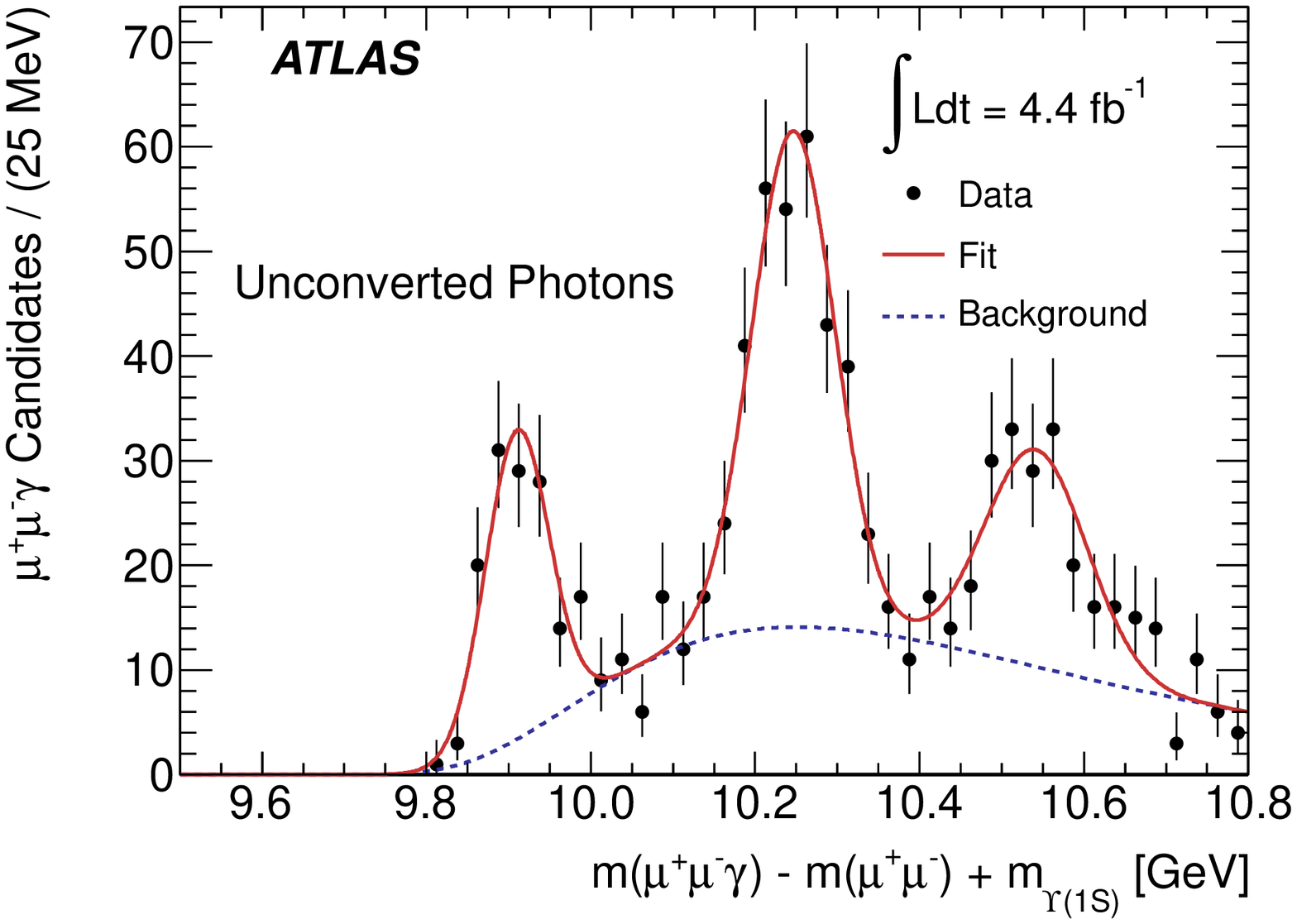}
  \label{fig.PaperChiBMass_a}
}
\subfigure[]{
  \includegraphics[width=0.5\textwidth]{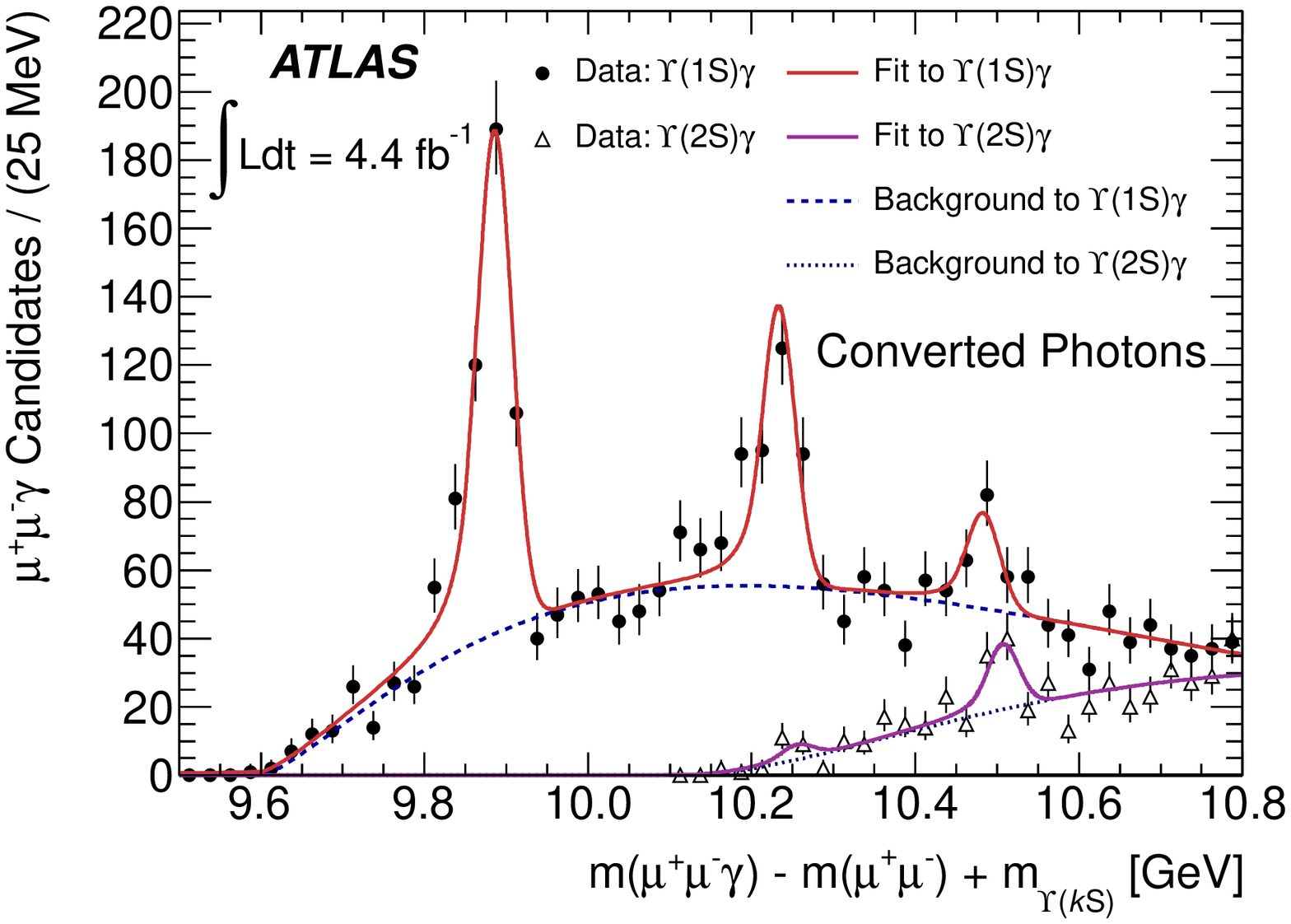}
  \label{fig.PaperChiBMass_b}
}
\caption{
(a) The mass distribution of $\chib\to\Upsilon(\textrm{1S})\gamma$ candidates for
unconverted photons reconstructed from energy deposits in the electromagnetic calorimeter ($\chi^{2}_\textrm{fit}/\textrm{d.o.f.}=0.85$). 
(b) The mass distributions of $\chib\to\Upsilon(k\textrm{S})\gamma$ ($k=1,2$) candidates formed using photons which have converted and been reconstructed in the ID
($\chi^{2}_\textrm{fit}/\textrm{d.o.f.}=1.3$). 
Data are shown before the correction for the energy loss from the photon conversion electrons due to bremsstrahlung and other processes.
The data for decays of $\chib\to\Upsilon(\textrm{1S})\gamma$ and $\chib\to\Upsilon(\textrm{2S})\gamma$ are plotted using circles and triangles respectively. 
Solid lines represent the total fit result for each mass window. The dashed lines represent the background components only.
}
\label{fig.PaperChiBMass}
\end{center}
\end{figure}
In addition to the expected peaks for $\chib(\textrm{1P,2P})\to\Upsilon{(1\textrm{S},2\textrm{S})}\gamma$, 
structures are observed at an invariant mass of approximately $10.5 \GeV$. These additional structures are interpreted as the radiative decays of the previously unobserved $\chibp3$ states, $\chibp3\rightarrow\Upsilon(\textrm{1S})\,\gamma$ and $\chibp3\rightarrow\Upsilon(\textrm{2S}) \,\gamma$. 

Separate fits are performed to the $\tilde{m}_{k}$ distributions of the selected $\mumu\gamma$ candidates reconstructed from converted and unconverted photons to extract mass information from the observed $\chibp3$ signals. 
The higher threshold for unconverted photons ($2.5\GeV$, versus $1\GeV$ for converted photons) 
prevents the reconstruction of the soft photons from $\chib(\textrm{2P,3P})$ decays into $\Upsilon(\textrm{2S})$.

An unbinned extended maximum likelihood fit is performed to the $\tilde{m}_{1} = \Delta m + m_{\Upsilon{(\textrm{1S})}}$ distribution of the selected unconverted $\mumu\gamma$ candidates. The three peaks in the distribution are each modeled by a Gaussian probability density function (pdf) with independent normalization parameter, $N_{n}$, mean value, $\overline{m}_{n}$, and width parameter, $\sigma_{n}$. The background distribution is parameterized by the pdf $N_{B}\cdot\exp{\left(A  \cdot  {\Delta m} + B \cdot  {\Delta m}^{-2} \right)}$ 
where $N_{B}$, $A$, and $B$ are all free parameters. 
The three mean values $\overline{m}_{n=1,2,3}$ determined by the fit are shown in Table~\ref{tab.FitNumbers}. The mean value $\overline{m}_{3}$ is an estimate 
of the mass barycenter of the observed $\chibp3$ signal.

\begin{table*}
\centering
  \caption{The fitted mass of the $\chib ({n\textrm{P}})$ signals for both converted and unconverted photons. 
    The systematic uncertainty on the mass of candidates reconstructed with unconverted photons is determined in the same way for all three states.
    Also included are theoretical predictions~\cite{Chib3P,Chib3P_splitting} for the spin-averaged masses of the $\chib$ states.
  }
\begin{ruledtabular}
\begin{tabular}{ccc c cc r@{ $\pm$}r@{ $\pm$}l c r@{ $\pm$}l@{ $\pm$}l}
	 \multirow{2}{*}{State} && \multicolumn{3}{c}{Model predictions}                 && \multicolumn{7}{c}{Fitted masses [MeV]}  \\ 
                                && \multicolumn{3}{c}{\cite{Chib3P,Chib3P_splitting} [MeV]} && \multicolumn{3}{c}{Unconverted Photons} && \multicolumn{3}{c}{Converted Photons}  \\ 
                                [-0.3 cm]\\\hline \\[-0.3 cm]
	 $\chibp1$ &&& \multicolumn{1}{r}{\hspace{1.5cm}$9900$}&                                      && $9910$&$6 \stat$&$11 \syst$        && \multicolumn{3}{l}{Fixed to $\chi_{b1}=$\hspace{0.25cm}$9892.78$ \& $\chi_{b2}=$\hspace{0.25cm}$9912.21$~\cite{PDG}}\\ 
	 
	 $\chibp2$ &&& \multicolumn{1}{r}{$10260$}&                                     && $10246$&$5 \stat$&$18 \syst$      && \multicolumn{3}{l}{Fixed to $\chi_{b1}=10255.46$ \& $\chi_{b2}=10268.65$~\cite{PDG}}\\ 
	 
	 $\chibp3$ &&& \multicolumn{1}{r}{$10525$}&                                     && $10541$&$11 \stat$&$30 \syst$     && $10530$&$5 \stat $&$9 \syst$  \\ 
    \end{tabular}     \label{tab.FitNumbers}
\end{ruledtabular}
\end{table*}

Likewise, the $\tilde{m}_{1} = \Delta m + m_{\Upsilon(\textrm{1S})}$ and $\tilde{m}_{2} = \Delta m + m_{\Upsilon(\textrm{2S})}$ distributions for the sample of $\mumu\gamma$ candidates reconstructed from converted photons are fitted using an unbinned extended maximum likelihood method.
A simultaneous fit is performed on the $\tilde{m}_{1}$ and $\tilde{m}_{2}$ distributions for the $\chib{(n\textrm{P})}\to\Upsilon(\textrm{1S})\gamma$ (for $n=1,2,3$) and $\chib{(n\textrm{P})}\to\Upsilon(\textrm{2S})\gamma$ (for $n=2,3$ only) signals, with the distributions modeled by three signal components (two of which are shared between the  $\Upsilon(\textrm{1S})$ and $\Upsilon(\textrm{2S})$ distributions) and two background distributions.

In the $\Delta m$ distribution for the converted photon candidates the typical mass
resolution is found to be in the range $16-20 \MeV$, of similar magnitude to the hyperfine
splittings, motivating the need for multiple signal components for each of 
the $\chib{(n\textrm{P})}$ peaks. For $n=1,2$, the radiative branching fractions of the $J=0$ states are suppressed with respect to the $J=1,2$ states~\cite{PDG} and therefore a $J=0$ component is not included in the fit. Similar behavior is assumed for the $n=3$ case. Each of the three peaks ($n=1,2,3$) is therefore parameterized by a doublet of Crystal Ball (CB) \cite{CrystalBall} functions (corresponding to $J=1,2$ states) with resolution $\sigma$ and radiative tail parameters
common to all peaks. For $n=1$ and $n=2$, the peak mass values and hyperfine splittings are fixed to the world averages~\cite{PDG} for the respective $\chib$ states (see Table~\ref{tab.FitNumbers}). For $n=3$, the hyperfine mass splitting is fixed to the theoretically predicted value of $12 \MeV$~\cite{Chib3P_splitting}, while the average mass is left as a free parameter. The unknown relative normalization of the $J=1$ and $J=2$ CB peaks is taken to be equal and treated as a systematic uncertainty (for all doublets) for the baseline fit.

In order to take into account energy loss from the photon conversion electrons due to bremsstrahlung and other processes, the measured values of $\Delta m$ in the
$\tilde{m}_{1}$ and $\tilde{m}_{2}$ distributions are scaled by a common parameter $\lambda=0.961\pm 0.003$, which determines the energy scale 
and is derived from the fit to the $\chi_{b}(\textrm{1P},\textrm{2P})$ signals. 
The background components of the $\Delta m$ distributions for the $\Upsilon(\textrm{1S})\gamma$ and $\Upsilon(\textrm{2S})\gamma$ final states are each modeled by the 
pdf $N^{k}_{B}\cdot\left( \Delta m - q^0_k\right)^{A_{k}}\cdot\exp{\left[B_{k}\cdot \left(\Delta m - q^0_k\right)\right]}$ for $\Delta m>q^0_k$, and zero otherwise, 
where $N^{k}_{B}$, $q^0_k$, $A_{k}$, and $B_{k}$ ($k=1,2$) are all free parameters. 
The mean value $\overline{m}_{3}$ determined by the fit is shown in Table~\ref{tab.FitNumbers}.

In the fit using unconverted photons, the signal is refitted using an alternative (two Gaussians) model for each of the three $\chib$ states, resulting in a negligible change in the
peak positions.
Alternative fits to the background are also used, either including constraints on the $\Delta m$ distribution using di-muon pairs from the 
low-mass ($8.0 \GeV < m_{\mu\mu} < 8.8 \GeV$) sideband or different background pdfs.
The systematic uncertainty on the $\chibp3$ mass barycenter from modeling of the background distribution is determined to be $\pm 21 \MeV$.
The systematic uncertainty associated with the unconverted photon energy scale is estimated to be $\pm 2\%$ on the $\Delta m$ position, corresponding to a systematic
uncertainty on $\overline{m}_{3}$ of $\pm 22 \MeV$. The uncertainties due to background modeling and photon energy scale comprise the dominant sources of systematic uncertainty.

For the fit using converted photons, alternative signal and background models are compared, as well as releasing various constraints in the fit model. 
The unknown relative normalizations of the $J=1$ and $J=2$ CB peaks are varied both coherently and incoherently between the 1P, 2P and 3P doublets by $\pm0.25$,
resulting in a maximum variation in $\overline{m}_{3}$ of $\pm5 \MeV$. 
Smaller variations are obtained if the common value of the relative normalization is allowed to be determined freely by the fit to the three doublets.
Background modeling variations, decoupled fits to the $\tilde{m}_{1}$ and $\tilde{m}_{2}$ distributions,
and individually released constraints on the mass position of the $n=1,2$ doublets 
each result in deviations of the order of $\pm5 \MeV$ or smaller. Furthermore, if the constraints on the masses of the $n=1,2$ peaks are released, the values obtained from the fit are consistent with expectations~\cite{PDG}, within statistical errors and uncertainty in the relative contributions from $J=1$ and $J=2$ states.
The effect of symmetrizing the $\Upsilon(\textrm{2S})$ mass window is studied and found to have negligible effect on the fitted $\chib$ masses while
increasing background contamination.
The resulting shifts in $\overline{m}_{3}$ for these independent variations are added in quadrature to provide an estimate of the systematic uncertainty.

The $\chibp3$ signal significance is assessed from $\log(L_\textrm{max}/L_0)$, where $L_\textrm{max}$ and $L_0$ are the likelihood values from the 
nominal fit and from a fit with no $\chibp3$ signal included, respectively. The fit is repeated with each of the systematic variations
in the model, as discussed above, and the likelihood ratio re-evaluated. The significance of the
$\chibp3$ signal is found to be in excess of six standard deviations in each of the unconverted and converted photon selections independently.

The mass barycenter for the $\chibp3$ signal, determined from the fit using unconverted photon candidates is:
$$\overline{m}_{3} = 10.541\pm 0.011 \stat \pm 0.030 \syst \GeV  .$$
The mass barycenter for the $\chibp3$ signal, determined from the fit using converted photon candidates is:
$$\overline{m}_{3} = 10.530\pm 0.005 \stat \pm 0.009 \syst \GeV  .$$

The measured mass barycenters of the $\chibp1$, $\chibp2$ and $\chibp3$ systems are summarized in Table~\ref{tab.FitNumbers}.
The results of the converted and unconverted photon analyses for the $\chibp3$ are found to be compatible. 
Given the substantially smaller systematic uncertainties in the conversion measurement, 
the final mass determination for $\overline{m}_{3}$ is quoted solely on the basis of this analysis.

In conclusion, the production of the heavy quarkonium states $\chib{(n\textrm{P})}$ in proton-proton 
collisions at $\sqrt{s} = 7\TeV$ is observed through reconstruction of the radiative decay modes of
$\chib {(n\textrm{P})}\to\Upsilon(\textrm{1S,2S})\,\gamma$. 
Mass peaks corresponding to $\chib(\textrm{1P,2P})$ decays are observed, 
together with additional structures at higher mass, which are consistent with theoretical predictions for 
$\chibp3\rightarrow\Upsilon(\textrm{1S})\,\gamma$ and $\chibp3\rightarrow\Upsilon(\textrm{2S}) \,\gamma$.
These observations are interpreted as the $\chibp3$ multiplet, the mass barycenter of which is measured to be $10.530\pm 0.005 \stat \pm 0.009 \syst \GeV$.

\section{Acknowledgements}
We thank CERN for the very successful operation of the LHC, as well as the
support staff from our institutions without whom ATLAS could not be
operated efficiently.

We acknowledge the support of ANPCyT, Argentina; YerPhI, Armenia; ARC,
Australia; BMWF, Austria; ANAS, Azerbaijan; SSTC, Belarus; CNPq and FAPESP,
Brazil; NSERC, NRC and CFI, Canada; CERN; CONICYT, Chile; CAS, MOST and
NSFC, China; COLCIENCIAS, Colombia; MSMT CR, MPO CR and VSC CR, Czech
Republic; DNRF, DNSRC and Lundbeck Foundation, Denmark; ARTEMIS, European
Union; IN2P3-CNRS, CEA-DSM/IRFU, France; GNAS, Georgia; BMBF, DFG, HGF, MPG
and AvH Foundation, Germany; GSRT, Greece; ISF, MINERVA, GIF, DIP and
Benoziyo Center, Israel; INFN, Italy; MEXT and JSPS, Japan; CNRST, Morocco;
FOM and NWO, Netherlands; RCN, Norway; MNiSW, Poland; GRICES and FCT,
Portugal; MERYS (MECTS), Romania; MES of Russia and ROSATOM, Russian
Federation; JINR; MSTD, Serbia; MSSR, Slovakia; ARRS and MVZT, Slovenia;
DST/NRF, South Africa; MICINN, Spain; SRC and Wallenberg Foundation,
Sweden; SER, SNSF and Cantons of Bern and Geneva, Switzerland; NSC, Taiwan;
TAEK, Turkey; STFC, the Royal Society and Leverhulme Trust, United Kingdom;
DOE and NSF, United States of America.

The crucial computing support from all WLCG partners is acknowledged
gratefully, in particular from CERN and the ATLAS Tier-1 facilities at
TRIUMF (Canada), NDGF (Denmark, Norway, Sweden), CC-IN2P3 (France),
KIT/GridKA (Germany), INFN-CNAF (Italy), NL-T1 (Netherlands), PIC (Spain),
ASGC (Taiwan), RAL (UK) and BNL (USA) and in the Tier-2 facilities
worldwide.

\providecommand{\href}[2]{#2}\begingroup\raggedright\endgroup

\onecolumngrid
\clearpage
\onecolumngrid
\begin{flushleft}
{\Large The ATLAS Collaboration}

\bigskip

G.~Aad$^{\rm 48}$,
B.~Abbott$^{\rm 110}$,
J.~Abdallah$^{\rm 11}$,
A.A.~Abdelalim$^{\rm 49}$,
A.~Abdesselam$^{\rm 117}$,
O.~Abdinov$^{\rm 10}$,
B.~Abi$^{\rm 111}$,
M.~Abolins$^{\rm 87}$,
O.S.~AbouZeid$^{\rm 157}$,
H.~Abramowicz$^{\rm 152}$
% ...
\bigskip

% ...
$^{10}$ Institute of Physics, Azerbaijan Academy of Sciences, Baku, Azerbaijan\\
$^{11}$ Institut de F\'isica d'Altes Energies and Departament de F\'isica de la Universitat Aut\`onoma  de Barcelona and ICREA, Barcelona, Spain\\
% ...
$^{48}$ Fakult\"{a}t f\"{u}r Mathematik und Physik, Albert-Ludwigs-Universit\"{a}t, Freiburg i.Br., Germany\\
$^{49}$ Section de Physique, Universit\'e de Gen\`eve, Geneva, Switzerland\\
% ...
$^{87}$ Department of Physics and Astronomy, Michigan State University, East Lansing MI, United States of America\\
% ...
$^{110}$ Homer L. Dodge Department of Physics and Astronomy, University of Oklahoma, Norman OK, United States of America\\
$^{111}$ Department of Physics, Oklahoma State University, Stillwater OK, United States of America\\
% ...
$^{117}$ Department of Physics, Oxford University, Oxford, United Kingdom\\
% ...
$^{152}$ Raymond and Beverly Sackler School of Physics and Astronomy, Tel Aviv University, Tel Aviv, Israel\\
% ...
$^{157}$ Department of Physics, University of Toronto, Toronto ON, Canada\\
% ...
\end{flushleft}


\end{document}
